# Dynamics of Two-Level System Interacting with Random Classical Field

G. B. Lesovik, A. V. Lebedev*, and A. O. Imambekov

*Landau Institute for Theoretical Physics, Russian Academy of Sciences, ul. Kosygina 2, Moscow, 117334 Russia*
*\* e-mail: alebedev@itp.ac.ru*

The dynamics of a particle interacting with random classical field in a two-well potential is studied by the functional integration method. The probability of particle localization in either of the wells is studied in detail. Certain field-averaged correlation functions for quantum-mechanical probabilities and the distribution function for the probabilities of final states (which can be considered as random variables in the presence of a random field) are calculated. The calculated correlators are used to discuss the dependence of the final state on the initial state. One of the main results of this work is that, although the off-diagonal elements of density matrix disappear with time, a particle in the system is localized incompletely (wave-packet reduction does not occur), and the distribution function for the probability of finding particle in one of the wells is a constant at infinite time.

The standard interpretation of quantum mechanics includes the concept of the wave-packet reduction (WPR) in the act of measurement [1]. The question of the WPR mechanism either remains beyond the theory or is postulated. In [2], it was suggested that the reservoir be regarded as a WPR source. In this work, we used a two-level system as an example to study in detail the influence of the reservoir degrees of freedom (which were modeled by a random classical field) on the localization process.

The decoherence concept, which was developed over the past 20 years, leaves the WPR problem open [3]. The researchers are typically interested in the evolution of the density matrix of a quantum system interacting with the surrounding medium. With such an approach, much information on the dynamics of the system is lost after averaging over the medium degrees of freedom, rendering the WPR problem unresolved. In [2], it was suggested that the WPR be considered in terms of quantities other than the density matrix.

In this work, we used the simple two-level system as an example to demonstrate the method for and the results of calculating the quantities of this sort.

Let us consider a particle interacting with a random classical field in a double-well potential. The problem of interest is as follows: the particle is held in the left well until $t = 0$, whereupon it is left go. In what follows, we are interested in the quantities averaged over the medium degrees of freedom. Since, after averaging, the probability of finding particle in the left well does not carry full information about the dynamics [2], we will be interested not only in the medium-averaged probability $\langle P_{L \to L}(t) \rangle$ of finding particle in the left well (the first and second indices denote the initial and final states, respectively), but also in the correlators of the form

$$\langle P_{L \to L}(t) P_{L \to R}(t) \rangle, \qquad (1)$$

where the parentheses stand for the averaging over the degrees of freedom of the reservoir. The necessity for calculating the correlators of this type follows from the fact that the probabilities given by the density matrix are the result of both quantum-mechanical averaging and averaging over the medium degrees of freedom. For instance, $\langle P_{L \to L}(t = \infty) \rangle$ can become equal to 1/2 by various ways. A situation is possible, for which, depending on the reservoir state, either $P_{L \to L}(t = \infty) = 1$ or $P_{L \to L}(t = \infty) = 0$, while $P(P_{L \to L}(t)) = 1/2$ is obtained only after averaging. In this case, the WPR occurs in the model considered, and zero value of correlator (1) is an unambiguous evidence of this fact. In the more general case, there is a certain probability density $P_{L \to L}(t)$ of the reservoir states for which the particle occurs in the right well with the probability $P_{L \to L}(t)$. If all correlators of the form $\langle P_{L \to L}(t)^n \rangle$ are known, one can determine the quantity $P(P_{L \to L}(t))$.

Let us formulate the model in more detail. Assume that the wells are symmetric, so that the ground level is degenerate in the absence of tunneling. Under certain conditions [4], the Hilbert space of particle states can be thought of as being two-dimensional.

The states for which the particle coordinate takes the definite values $\pm q_0/2$ are chosen as the basis set. In the presence of tunneling, the Hamiltonian of a particle not interacting with the field has the form

$$H_0 = -\frac{1}{2}\hbar\Delta\sigma_x.$$

Here, $\sigma_x$ is the Pauli matrix, and the basis is chosen so that the eigenvalue $+1(-1)$ of the matrix $\sigma_z$ corresponds to the particle localized in the right(left) well. The interaction of the field with particle is taken into account by adding to the Hamiltonian the term $q\varphi(t)$ linear in field. In this model, the random field is determined by the external medium. We assume that the probability distribution for the field $\varphi(t)$ is Gaussian and the field correlator has the white-noise form

$$\langle\varphi(t_1)\varphi(t_2)\rangle = \delta(t_1 - t_2)\langle\varphi^2\rangle. \quad (2)$$

In this case, the averaging is carried out over the degrees of freedom of the surrounding medium, which induces the uncontrolled field deviations from zero. The Hamiltonian of the particle is

$$H(t) = -\frac{1}{2}\hbar\Delta\sigma_x + \frac{q_0\varphi(t)}{2}\sigma_z. \quad (3)$$

This model is equivalent to the spin 1/2 in a magnetic field, whose $x$ component is fixed, while the component along the $z$ axis is random.

Making use of the influence functional [5], one can write the probability $\langle P_{L\to L}(t)\rangle$ as a double functional path integral

$$(\langle P_{L\to L}(t)\rangle = \iint Dq_1(\tau 1)Dq_2(\tau 2)A[q_1(\tau 1)]) \\ \times A^*[q_2(\tau 2)]F[q_1(\tau 1), q_2(\tau 2)], \quad (4)$$

where the integral is taken over all paths for which $q_1(0) = q_2(0) = q_1(t) = q_2(t) = -q_0/2$, and $A[q(\tau)]$ is the amplitude for the path $q(\tau)$ in the absence of random field; $F[q_1(\tau 1), q_2(\tau 2)]$ is the influence functional that is equal, for the random Gaussian potential with correlator (2), to [5]

$$F[q_1(\tau 1), q_2(\tau 2)] \\ = \left\langle \exp\left\{-\frac{i}{\hbar}\int_0^t \varphi(\tau)(q_1(\tau) - q_2(\tau))d\tau\right\}\right\rangle \quad (5) \\ = \exp\left\{-\frac{\langle\varphi^2\rangle}{2\hbar^2}\int_0^t (q_1(\tau) - q_2(\tau))^2 d\tau\right\}.$$

At any instant of time, the pair of paths $[q_1, q_2]$ is in one of the four states $[-, -]$, $[-, +]$, $[+, -]$, and $[+, +]$ states, which will be denoted as $A$, $B$, $C$, and $D$. Introduce the notation $\xi(t) = q_0^{-1}(q_1(t) - q_2(t))$. Then the influence functional is recast as

$$F[q_1(\tau 1), q_2(\tau 2)] = \exp\left\{-\Gamma\int_0^T \xi(t)^2 dt\right\}, \quad (6)$$

$$\Gamma = q_0^2\langle\varphi^2\rangle/2\hbar^2. \quad (7)$$

Following the formalism described in [4], we expand $\langle P_{L\to L}(t)\rangle$ in powers of $i\Delta/2$. This multiplier (except for a sign) appears at every jump between the wells. We will describe each state $[[q_1, q_2]$ at every instant of time as a four-dimensional vector $\mathbf{E}_i$, where $i = \{1, 2, 3, 4\}$ corresponds to the $\{A, B, C, D\}$ states. The matrix of possible jumps has the form (the sign corresponds to the sign of transition amplitude)

$$\Lambda = \begin{bmatrix} 0 & -1 & 1 & 0 \\ -1 & 0 & 0 & 1 \\ 1 & 0 & 0 & -1 \\ 0 & 1 & -1 & 0 \end{bmatrix}. \quad (8)$$

Let us introduce the matrix allowing for the path weights due to the influence functional,

$$\mathbf{U}(t) = \begin{bmatrix} 1 & 0 & 0 & 0 \\ 0 & e^{-\Gamma t} & 0 & 0 \\ 0 & 0 & e^{-\Gamma t} & 0 \\ 0 & 0 & 0 & 1 \end{bmatrix}, \quad (9)$$

Then $\langle P_{L\to L}(t)\rangle$ can be written as

$$\langle P_{L\to L}(t)\rangle = \mathbf{E}_1^T \\ \times \left[\sum_{n=0}^{\infty}\int_0^t dt_n \int_0^{t_n} dt_{n-1}\ldots\int_0^{t_2} dt_1 \mathbf{U}(t-t_n)\mathbf{S}\ldots\mathbf{S}\mathbf{U}(t_1)\right] \times \mathbf{E}_1, \quad (10)$$

where $\mathbf{S} = i\Delta/2\Lambda$; $t_1, \ldots, t_n$ are the hopping times; and the vector $\mathbf{E}_1 = \{1, 0, 0, 0\}^T$ corresponds to the state $A$. Applying the Laplace transform to $\langle P_{L\to L}(t)\rangle$,

$$\langle P_{L\to L}(\lambda)\rangle = \int_0^{\infty}\langle P_{L\to L}(t)\rangle e^{-\lambda t}dt.$$

and changing the integration variables in Eq. (10), one obtains

$$\langle P_{L\to L}(\lambda)\rangle = \mathbf{E}_1^T \times \left[\sum_{n=0}^{\infty}[\mathbf{U}(\lambda)\mathbf{S}]^n\right]\mathbf{U}(\lambda)\times \mathbf{E}_1 \\ = \mathbf{E}_1^T \times [\mathbf{U}^{-1}(\lambda) - \mathbf{S}] \times \mathbf{E}_1, \quad (11)$$

where $\mathbf{U}(\lambda)$ is the Laplace transform of $\mathbf{U}(t)$:

$$\mathbf{U}(\lambda) = \begin{bmatrix} \lambda^{-1} & 0 & 0 & 0 \\ 0 & (\lambda+\Gamma)^{-1} & 0 & 0 \\ 0 & 0 & (\lambda+\Gamma)^{-1} & 0 \\ 0 & 0 & 0 & \lambda^{-1} \end{bmatrix}. \quad (12)$$

Therefore, the calculation of $\langle P_{L \to L}(t) \rangle$ amounts to the evaluation of the matrix element of the inverse of a $4 \times 4$ matrix following by taking the inverse Laplace transform. The result is

$$\langle P_{L \to L}(\lambda) \rangle = \frac{1}{2\lambda} \frac{2\lambda^2 + 2\lambda\Gamma + \Delta^2}{\lambda^2 + \Gamma\lambda + \Delta^2}.$$

To find $\langle P_{L \to L}(t=\infty) \rangle$, it suffices to know only the residue of $\langle P_{L \to L}(\lambda) \rangle$ at $\lambda=0$, which is equal to the inverse Laplace transform at $t = \infty$. The quantity $\langle P_{L \to L}(t) \rangle$ can be exactly calculated to give

$$\frac{1}{2} + \frac{\exp\left\{\frac{1}{2}t(-\Gamma + \sqrt{\Gamma^2 - 4\Delta^2})\right\}(\Gamma + \sqrt{\Gamma^2 - 4\Delta^2})}{4\sqrt{\Gamma^2 - 4\Delta^2}}$$
$$+ \frac{\exp\left\{-\frac{1}{2}t(\Gamma + \sqrt{\Gamma^2 - 4\Delta^2})\right\}(-\Gamma + \sqrt{\Gamma^2 - 4\Delta^2})}{4\sqrt{\Gamma^2 - 4\Delta^2}}. \quad (13)$$

One can see that $\langle P_{L \to L}(t=\infty) \rangle = 1/2$ for the nonzero $\Gamma$; at $\Gamma = 2\Delta$, the damping oscillations give way to relaxation.

In the limit $\Gamma \gg \Delta$, two relaxation times appear in the system. One of them, $\tau_1 = \Gamma^{-1}$, is considerably shorter than the other, $\tau_2 = \Gamma/\Delta^2$.

The off-diagonal elements $\langle \Psi_L(t)^* \Psi_R(t) \rangle$ of density matrix can be calculated in a similar way. The result

$$\langle \Psi_L(t)^* \Psi_R(t) \rangle = \frac{i\Delta \exp\left\{-\frac{1}{2}t(\Gamma + \sqrt{\Gamma^2 - 4\Delta^2})\right\}}{2\sqrt{\Gamma^2 - 4\Delta^2}}$$
$$- \frac{i\Delta \exp\left\{\frac{1}{2}t(-\Gamma + \sqrt{\Gamma^2 - 4\Delta^2})\right\}}{2\sqrt{\Gamma^2 - 4\Delta^2}} \quad (14)$$

also has two characteristic times, with $\langle \Psi_L(\infty)^* \Psi_R(\infty) \rangle = 0$. In the limit $\Gamma \gg \Delta$, the maximum magnitude $\Delta/2\Gamma$ is reached in a time on the order of $\tau_1$.[1]

To calculate the quantity $\langle P_{L \to L}(t) P_{L \to L}(t) \rangle$, one can also use the formalism developed above. In this case, the path integral is taken over four trajectories, and the system state is described at every instant by a 16-dimensional vector. After introducing for each pair of trajectories their own variables $\xi_1(t) = q_0^{-1}(q_1(t) - q_2(t))$ and $\xi_2(t) = q_0^{-1}(q_3(t) - q_4(t))$, the four-point influence functional can be written, similar to Eq. (6), as

$$F[q_1(\tau 1), q_2(\tau 2), q_3(\tau 3), q_4(\tau 4)]$$
$$= \exp\left\{-\Gamma \int_0^t (\xi_1(\tau) + \xi_2(\tau))^2 d\tau\right\}. \quad (15)$$

Introducing, by analogy with Eqs. (8) and (12), the 16-dimensional matrices $\mathbf{S2}(t)$ and $\mathbf{U2}(t)$ (we do not give here their explicit form), one arrives at the formula analogous to Eq. (11):

$$\langle P_{L \to L}(t) P_{L \to L}(t) \rangle = \mathbf{E}_1^T \times [\mathbf{U2}^{-1}(\lambda) - \mathbf{S2}]^{-1} \times \mathbf{E}_1 \quad (16)$$

where the vector $\mathbf{E}_1$ corresponds to the state $[-,-,-,-]$ of the paths $[q_1, q_2, q_3, q_4]$.

Calculation gives

$$\langle P_{L \to L}^2(t) \rangle = \frac{1}{3\lambda} + \frac{\Gamma + \lambda}{2(\Delta^2 + \Gamma\lambda + \lambda^2)}$$
$$+ \frac{\Delta^2 + (\Gamma + \lambda)(4\Gamma + \lambda)}{6(4\Delta^2(3\Gamma + \lambda) + \lambda(\Gamma + \lambda)(4\Gamma + \lambda))}. \quad (17)$$

For the infinite time, one has $\langle P_{L \to L}(t=\infty)^2 \rangle = 1/3$. In the limit $\Gamma \gg \Delta$, this correlator also has two characteristic times, which are determined by the real parts of the poles of $\langle P_{L \to L}^2(t) \rangle$. To the first nonvanishing terms, the relaxation times in Eq. (17) are equal to $\tau_1$, $\tau_1$, $\tau_1/4$, $\tau_2$, and $\tau_2/3$.

---

[1] Note that, since the particle in our problem (or "spin" in the equivalent problem) interacts with the classical field, it is formally described by a pure density matrix. This is natural, because the particle cannot act on the classical field, so that the entangled quantum states do not appear. Nevertheless, from the practical viewpoint, the distinction between the presence of a "real" reservoir and the random classical field is insignificant, because, to prove that the particle is in a pure state, one must conduct a set of measurements. For example, to obtain the definite result in a single measurement, one must know exact data on the magnitude of fluctuating classical field $\varphi(t)$, whose monitoring at the exact "spin" location is highly conjectural. Moreover, the temporal dynamics of the probabilities $\langle P_{L \to L}(t) \rangle$ are identical in both cases.

Similar calculations give for the remaining second-order correlators

$$\langle P_{L \to R}(t = \infty)^2 \rangle = \frac{1}{3},$$
$$\langle P_{L \to L}(t = \infty) P_{L \to R}(t = \infty) \rangle = \frac{1}{6}. \quad (18)$$

Thus, although the off-diagonal elements of density matrix vanish with time, $\langle P_{L \to L}(t = \infty) P_{L \to R}(t = \infty) \rangle \neq 0$, so that the particle localization (wave-packet reduction) in the system does not occur.

It is straightforward to generalize the above computational procedure to the case of $\langle P_{L \to L}(t)^n \rangle$, although the sizes of corresponding matrices rapidly increase. Numerical computations show that

$$\langle P_{L \to L}(t = \infty)^3 \rangle = \frac{1}{4}, \quad \langle P_{L \to L}(t = \infty)^4 \rangle = \frac{1}{5}. \quad (19)$$

Thus, we assume (although we have not succeeded in obtaining the general proof) in our model that the probability density $P(P_{L \to L}(\infty))$ of the reservoir states for which the particle occurs in the left well with the probability $P_{L \to L}(\infty)$ in an infinite time is unity on the interval (0, 1). Indeed, in this case

$$\langle P_{L \to L}^n \rangle = \int_0^1 (P_{L \to L})^n P(P_{L \to L}) d(P_{L \to L}) = \frac{1}{n + 1}, \quad (20)$$

which is fully consistent with our previous results. The correlators of the form $\langle P_{L \to L}^n P_{L \to R}^m \rangle$ are equal to

$$\langle P_{L \to L}^n P_{L \to R}^m \rangle = \int_0^1 P^n (1 - P)^m dP = \frac{n! m!}{(n + m + 1)!}. \quad (21)$$

Using the symmetry of matrices in Eq. (11), one can establish the following symmetry about the permutation of the initial and final states:

$$\langle P_{L \to L}^n P_{L \to R}^m \rangle = \langle P_{L \to L}^n P_{R \to L}^m \rangle. \quad (22)$$

Let us now consider how the final state depends on the initial state if the latter has the form

$$|S\rangle = a|\Psi_L\rangle + b|\Psi_R\rangle, \quad |a|^2 + |b|^2 = 1. \quad (23)$$

Using Eqs. (21) and (23), one obtains

$$\langle P_{S \to L}^n \rangle \quad (24)$$
$$= \langle (a^* \langle \Psi_{L \to L}| + b^* \langle \Psi_{R \to L}|)^n (a|\Psi_{L \to L}\rangle + b|\Psi_{R \to L}\rangle)^n \rangle$$
$$= (|a|^2 + |b|^2)^n / (n + 1) = 1/(n + 1).$$

Therefore, in the model considered, the distribution function for the probabilities in the final state is independent of the initial state after a very long time.

Nevertheless, the final state depends on the initial state in every *particular* case. The sensitivity to the initial state can be determined by calculating the following correlator:

$$\langle (P_{S \to L} - P_{S' \to L})^2 \rangle = \frac{|ab' - a'b|^2}{3}, \quad (25)$$

where $|S'\rangle = a'|\Psi_L\rangle + b'|\Psi_R\rangle$.

Consider, as an example, the case for which $a = b = 1/\sqrt{2}$ for one initial (ground) state $|S\rangle$ and $a' = b' = 1/\sqrt{2}$ for the other (excited) state $|S'\rangle$. Then $\langle (P_{S \to L} - P_{S' \to L})^2 \rangle = 1/3$. Note that the value $1/\sqrt{3}$ obtained for the mean difference in the final probabilities is larger than the average probability $1/2$.

We are now in position to discuss the sensitivity of the final state to the variations in external field. We formulate the problem in terms of a spin in an external magnetic field. Consider the correlator of the form $\langle (P_L\{H(t)\} - P_L\{H(t) + \delta H_z(t)\})^2 \rangle$. In the general form, the problem is complicated by the fact that the Hamiltonian $\mathbf{H}(t)\boldsymbol{\sigma}$ does not commute with itself at different instants of time. However, we can consider a particular case of the field $\delta H_z(t)$ acting during a short time interval (such that the spin rotation about the $x$ axis can be ignored because of the smallness of $H_x\sigma_x$ at the very beginning of state evolution. In this case, the problem reduces to the previous problem if the state $|S'\rangle = a'|\Psi_L\rangle + b'|\Psi_R\rangle$ is defined as

$$|S'\rangle = \exp\left[\frac{i}{\hbar}\int dt \delta H_z(t)\sigma_z\right][a|\Psi_L\rangle + b|\Psi_R\rangle]. \quad (26)$$

The new state $(a', b')$ is determined by the relative phase incursion $2\delta\Phi = (2/\hbar)\int \delta H_z(t)dt$ for $a$ and $b$. If the field pulse is applied at nonzero time, we can reformulate the problem starting with a certain fixed state $(a_0, b_0)$ at time $t_0$. In this case, one obtains after averaging

$$\langle (P_L\{H(t)\} - P_L\{H(t) + \delta H_z(t)\})^2 \rangle_{t = \infty}$$
$$= \langle (P_L(t_0) P_R(t_0)) \rangle \frac{4 \sin^2(\delta\Phi)}{3}. \quad (27)$$

We note in conclusion that, in our opinion, the localization in a double-well potential will occur at long times, if a quantum reservoir is added to the classical reservoir. Qualitatively, this process can be imagined as follows: at every instant of time, the energy levels in the wells are different due to the classical field, the difference being sufficiently large for the instantaneous eigenstates of the Hamiltonian to be localized in either of the wells, while the transition to the lowest (localized) state occurs due to the photon emission.

We thank G.E. Volovik for helpful discussion. This work was supported by the Public Foundation for Assisting in Russian Science, the Russian Foundation for Basic Research (project no. 00-02-16617), the Ministry of Science (project "Physical fundamentals of quantum calculations"), the Dutch Scientific Foundation (grant for the collaboration with Russia), and the Swiss Scientific Foundation.